\documentclass[prd,superscriptaddress,onecolumn,showpacs,11pt,msmath,preprintnumbers,showkeys]{revtex4}
\usepackage{wrapfig,rotating}
\usepackage{graphicx,epsfig,color}



\usepackage{graphicx}
\usepackage{dcolumn}
\usepackage{bm}

\def\beq{\begin{equation}}
\def\eeq{\end{equation}}
\def\beeq{\begin{eqnarray}}
\def\eeeq{\end{eqnarray}}

\newcommand \Pomeron {I\!\!P}

\def\effs {$\sigma_{\textrm{\tiny eff}}\,$}
\def\2GPD{$_2\mbox{GPD}$}

\def\oo{$1 \otimes 1$}
\def\12{$1\otimes 2$}
\def\22{$2 \otimes 2$}
\def\eff{{\mbox{\scriptsize eff}}}

\def\Qsep{Q_{\mbox{\rm\scriptsize sep}}}
\def\Qsep2{Q^2_{\mbox{\rm\scriptsize sep}}}

\begin{document}
 \title{Open charm production in Double Parton Scattering processes in the forward kinematics}
 \author{B.\ Blok$^{1}$,
M. Strikman$^2$ \\[2mm] \normalsize $^1$ Department of Physics, Technion -- Israel Institute of Technology,
 Haifa, Israel\\
 \normalsize $^2$Physics Department, Pennsylvania State University, University Park,USA}
 \begin{abstract}
We calculate the rate of double open charm production in the forward kinematics studied recently in the  LHCb experiment. We find that  the mean field approximation for the double parton GPD (Generalized parton distributions),
which neglects parton - parton correlations,
underestimates the rate by a factor of two. The enhancement due to the  perturbative QCD correlation     \12 mechanism which explains the rate of  double parton interactions at  the central rapidities is found to explain
60 $\div$  80 \%
 of the discrepancy.
 We argue that the nonperturbative fluctuations leading to non-factorized
 (correlated)
  contributions to the initial conditions
for the DGLAP collinear evolution of  the double parton GPD  play an important role in this kinematics.  Combined, the two correlation mechanisms provide  a good description of the rate of double charm production reported by the LHCb. We also give predictions for the   variation of the \effs (i.e. the ratio of double and square of single inclusive rates) in the discussed kinematics as a function of $p_t$.  The account for two correlation  mechanisms strongly reduces sensitivity of the results to the starting point of the QCD evolution.

 \end{abstract}

   \maketitle
 \thispagestyle{empty}

 \vfill

\section{Introduction.}
\par It is widely realized now that hard {\em Multiple Parton Interactions}\/ (MPI)  play an important role in the description of inelastic proton-proton ($pp$) collisions at the LHC energies where MPIs occur with probability of the order one in  typical inelastic collisions.

Hence after years of relatively sparse theoretical activities after pioneering papers of the eighties \cite{TreleaniPaver82,mufti} studies of the MPI became a field of very active theoretical research,  see e.g. \cite{stirling,BDFS1,Diehl2,stirling1,BDFS2,Diehl,BDFS3,BDFS4,Gauntnew,Gauntdiehl,mpi2014,mpi2015,BG1,BG2} and references therein.
\par Also,
in the past several years  a number of Double Parton Scattering (DPS) measurements in different channels in the central rapidity  kinematics were carried out
\cite{tevatron1,tevatron2,tevatron3,cms1,atlas,cms2}, while many Monte Carlo (MC)  event generators now incorporate MPIs.

\par The recent discovery by the LHCb of the double charm DPS production attracted a lot of attention since it expands the study of
multiparton dynamics  into a new kinematics region of large rapidities \cite{Belyaev,LHCb,LHCb1,LHCb2}, and since the background from the leading twist processes is very strongly suppressed in this kinematics \cite{Likhoded,shurek,shurek1}.

The LHCb data are  available  for the  $J/\psi$ DPS production: $J/\psi-D\bar D$  and for the DPS production of two $D\bar D$ pairs.
According to the LHCb
experiment results, the DPS rate
 in the studied  kinematics,
 which is customarily parameterized by 1/\effs \, is    {\it
practically  the same for all channels} and
\effs $\sim$ 20 mb (see Fig. 10 in \cite{LHCb}).
The observed
universality of \effs \,  is consistent with expectations  of the approximation outlined below.
Here, as usual, \effs is defined as
\beq
\sigma_{\rm eff}=\sigma_1\sigma_2/\sigma_4
\eeq
where $\sigma_{1,2}$ are cross sections of elementary $2\rightarrow 2$ processes and $\sigma_4$ is a cross section
of a process $pp\rightarrow 1+2$ final state.
We will  focus on
production of two $D\bar D$ pairs since the data for  this channel have
 the  smallest errors \cite{vogt}. Also,  more complicated mechanisms than
  the  $gg \to J/\psi +X$ process may contribute in the case of
 $J/\psi$  production, i.e. $ggg\rightarrow J/\psi$ (see e.g. \cite{motyka} for a recent discussion).

It was pointed out starting with \cite{Frankfurt,Frankfurt2,BDFS1}
that the rate of DPS calculated under assumption that partons in nucleons are uncorrelated  (and using information about the gluon
GPDs  available from the analysis \cite{Frankfurt,Frankfurt2}
 of the HERA data) is too  low to explain the data.
It was pointed out in
\cite{BDFS4,Gauntnew,BG1,BG2,BO} that correlations generated in the course of the DGLAP evolution --  \12 mechanism -- explain the DPS
 rates in the central rapidity region  \cite{BDFS4,Gauntnew,BG1,BG2,BO}
 provided the starting scale for the QCD evolution -- $Q_0^2= 0.5 \div 1 {\rm GeV}^2$ is chosen.
  The remaining problem seems to be a strong enhancement of the processes involving  $J/\Psi$ production
\cite{D0,cms4} at $\sqrt{s}$ = 2TeV which does not show up in the LHCb data.

In this
letter we demonstrate  that the new LHCb data
\cite{Belyaev,LHCb,LHCb1,LHCb2} corresponding to the forward kinematics
can be explained by
taking into account two effects: buildup with increase of $Q^2$ of the perturbative correlations --
 the \12 mechanism, calculated using DGLAP formalism \cite{BDFS1,BDFS2,BDFS3,BDFS4}
and soft
 small $x$ parton - parton correlations in the nucleon wave function which result in a
non-factorized contribution to the initial conditions of the double parton GPD which  can be estimated using
information on diffraction in lepton / hadron -- nucleon scattering
following  the ideas first presented  in \cite{BDFS2}.

The paper
 is organized as follows.
   In section two we describe the kinematics of the LHCb experiment.  In the third  chapter we present that  the mean field approximation results for the rate of $DD$ production and demonstrate that they are a factor of two lower  than the data.  In section  four we present results for the  \12 mechanism contribution (see Fig. 2) to the cross section, In section five we discuss the Reggeon model based  estimate of the non-factorized contribution to the  initial conditions at  $Q_0^2\sim 0.5-1{\rm GeV}^2$, and its $Q^2$ evolution. In section six we present general formula for \effs \, combining the mean field,\12 and nonperturbative non-factorized contributions.
In section seven  we demonstrate that the simultaneous account of all three DPS
  mechanisms leads to the $\sigma_{eff}$ values consistent with the data.   The results are summarized in section eight.

\section{Kinematics of the LHCb study  of the double charm production}
\begin{figure}[htbp]
\begin{center}
\includegraphics[scale=0.30]{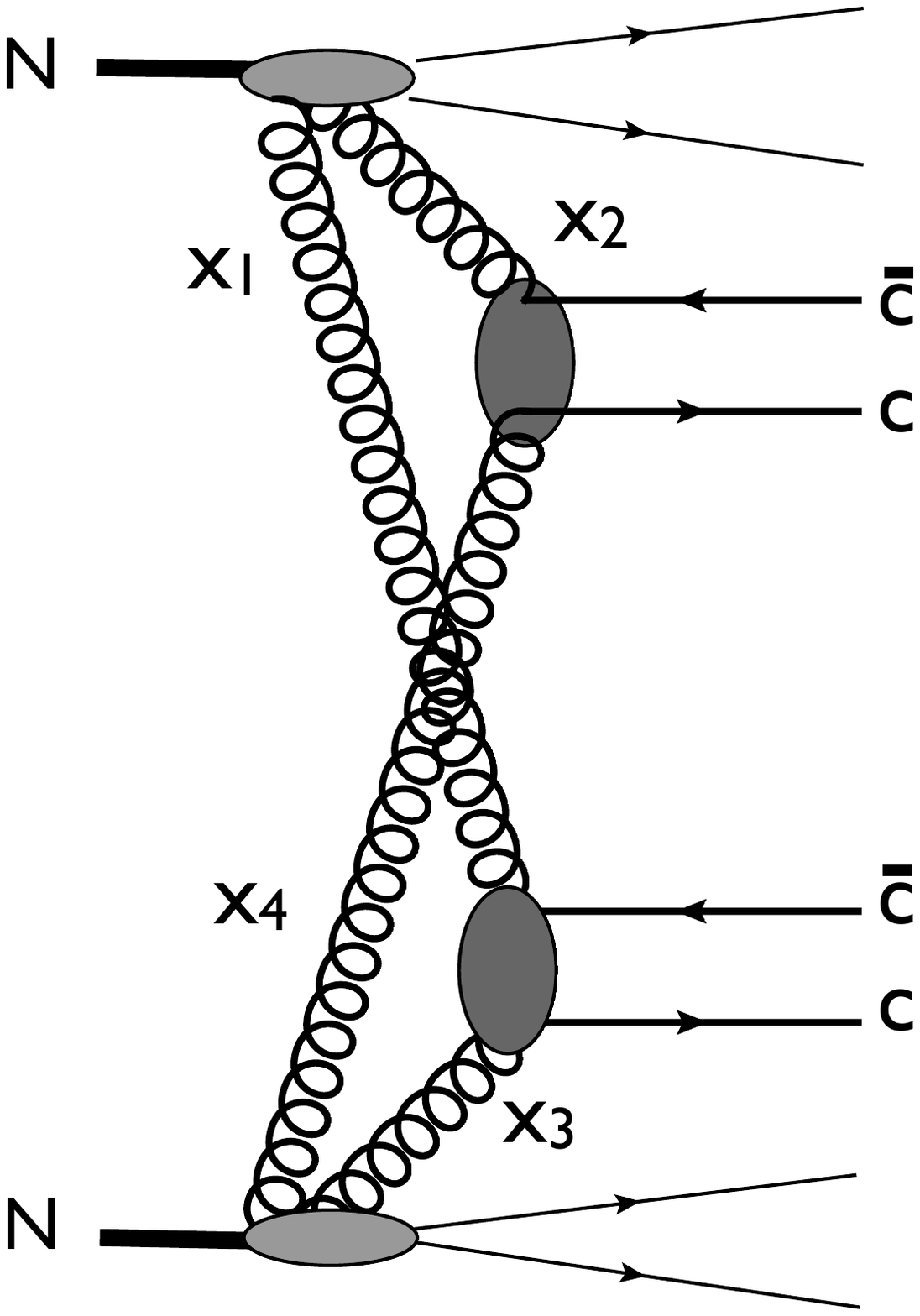}\hspace{-2cm}
\label{fig1k}
\caption{Kinematics of double charm production at LHCb.}
\end{center}
\end{figure}
\par So far LHCb experiment has presented results for  \effs \, integrated over  a significant range of rapidities and transverse momenta.
So in our analysis we will first perform  calculations for the typical LHCb kinematics and later on present the results for
the variation of \effs \, within the LHCb kinematic range which turns out to be pretty weak.

\par In the case of production of two D mesons, the main mechanism is production of  two pairs of $D\bar D$ mesons in two
 hard process (DPS)  (see Fig.1) with  two  D ($D_s$) mesons  originating from two
$D\bar D$  pairs. D-mesons are observed in the rapidity interval $y= 2 \div 5$,
 the average  rapidity interval  between $D$ and $\bar D$ meson is  of the order $\Delta y = 0.5$.
   A cutoff of $p_t\ge 3 {\rm GeV}$ was introduced in the DPS analysis leading to the average transverse momenta of the  $D$ mesons  of the order  of $p_t\sim 4$ GeV.
 Hence D-mesons
  are   created in the interaction of two gluons
  with virtualities
$Q^2\sim (2p_t^2+m_c^2)\sim 34$ GeV$^2$. The factor of two takes into account the fragmentation of $c\to D$ in which   D-mesons carry, in average,  $\sim 0.75$ fraction of the  jet momentum \cite{vogt} (see Fig.1).
\par The invariant mass squared  of the created D meson pair is $ x_1x_3s=4(p_t^2+0.5*m_c^2)\times 2\sim 136 $GeV$^2$, where the  factor two roughly accounts  for the fragmentation of the charmed quark into D meson, and spread of $D$ and $\bar D$ over
rapidities,
and $s=4.9\times 10^7$ GeV$^2$. The Bjorken $x$ of the  colliding gluon belonging to the proton moving in positive direction is determined from the condition  $x_3\sim p_t\exp(y)/(\sqrt{s}/2)\sim 0.01-0.02$, where $y\sim 3$ is the $ D$ meson rapidity.
The Bjorken $x$ of the gluon emitted by  the  nucleon moving in negative direction is given by
   $x_1\cdot x_3 s=136$ GeV$^2$  and is $0.0001-0.0002$.
   (In our notations $x_1, x_2$ correspond to small x gluons, while $x_3, x_4$ to large x ones.
The  effective cross section was  determined for several channels
\beq
\sigma_{\rm eff\,\, 2D 2\bar D}=\frac{\sigma_1\sigma_2}{\sigma_4(D\bar D)}\sim 20 {\rm mb},
\eeq
with a small  uncertainty for the channels with the highest statistics.

\par The important  advantage of these processes as compared to the processes experimentally studied before is that  in this kinematics the
SPS production of D-meson pairs is practically negligible \cite{Likhoded,shurek,shurek1}
and the dominant process  is the DPS production of $c\bar c$ pairs by gluons, thus permitting to use the methods developed in
\cite{BDFS1,BDFS2,BDFS3,BDFS4}.
\par Similar calculations can be carried out for double $b\bar b$ pair production and $b\bar b c \bar c $ pair production.
The only difference is that the corresponding transverse scale for b-pairs is $Q^2=m^2_b+1.5p_t^2\sim 50$ GeV$^2$ where we shall take $p_t\sim 4 $  GeV below as characteristic momenta. The corresponding invariant mass
squared
is of order $200 $ GeV$^2$
 and $x_1\sim 0.003,x_3\sim 0.014$.
\section{Mean field approximation estimate
of \effs}
\begin{figure}[htbp]
\begin{center}
\includegraphics[scale=0.35]{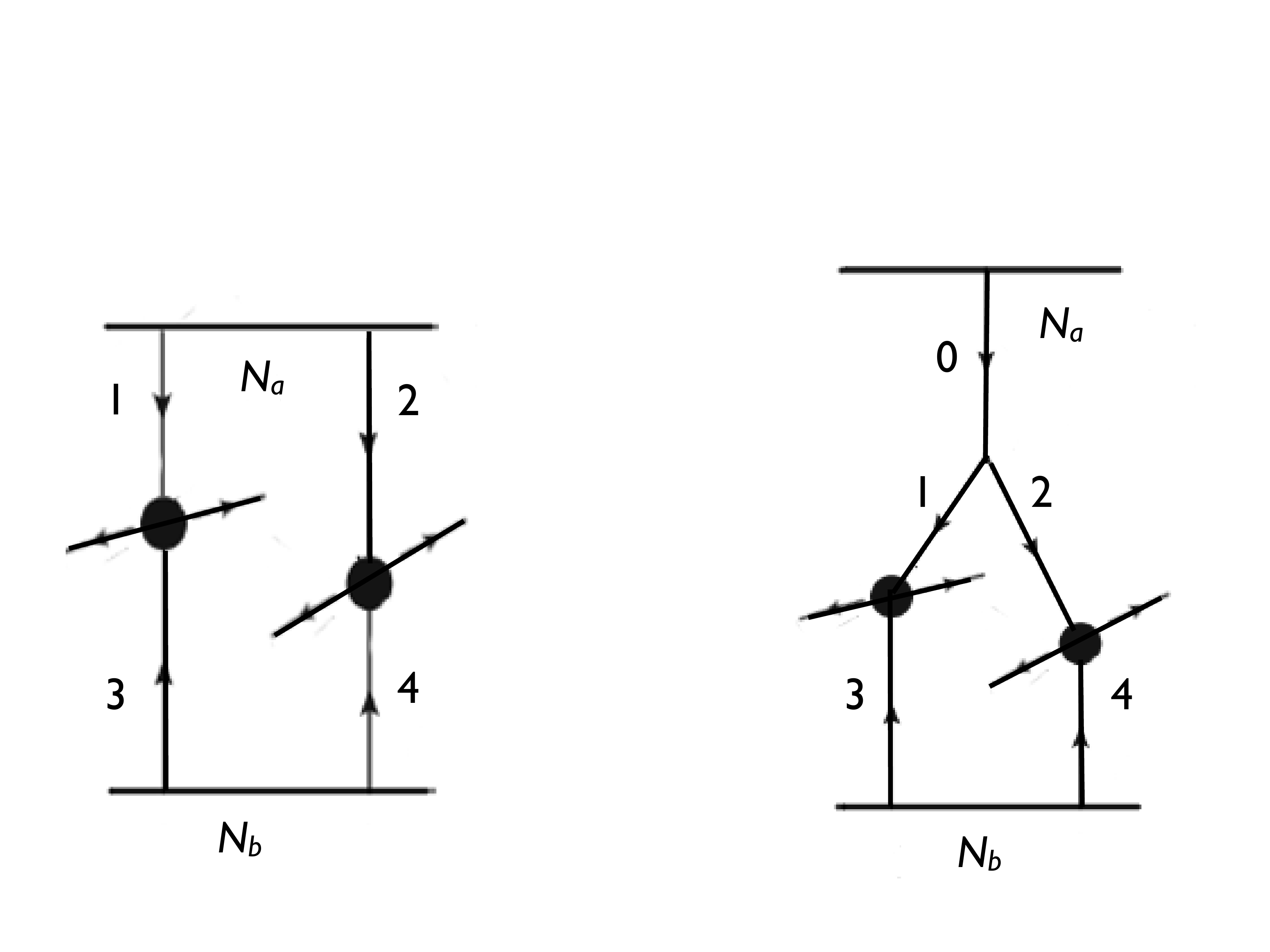}
\label{fig1a}
\caption{Sketch of the two considered DPS mechanisms: \22 (left) and \12 (right) mechanism.}
\end{center}
\end{figure}

 Recall that in the mean field approach  (see  Fig.2 left ) double  parton GPDs, describing the DPS, are
\beq
_2D(x_1,x_2,Q_1^2,Q_2^2,\Delta))= _1D(x_1,Q_1^2,\Delta_1) \cdot  _1D(x_2,Q_2^2,\Delta_2),\label{fac}
\eeq
where the one particle GPDs $_1D$ are  known from the analyses \cite{Frankfurt,Frankfurt1} of exclusive $J/\Psi$ photoproduction at HERA.
They  are parametrized as
\beq
D_1(x,Q^2,\Delta)=D(x,Q^2) F_{2g}(\Delta,x).\label{slon1}
\eeq
Here $D(x,Q^2)$
is the conventional gluon PDF of the nucleon, and $F_{2g}(\Delta,x)$ is the two gluon nucleon form factor.
The effective cross section \effs \, is then given by
\beq
1/\sigma_{eff}=\int \frac{d^2\Delta}{(2\pi)^2}F^4_{2g}(\Delta).
\eeq
We shall use exponential parametrization \cite{Frankfurt1}
\beq
F_{2g}(\Delta,x)=\exp(-B_g(x)\Delta^2/2),
\label{d1a}
\eeq
where $B_g$($x$)= $B_0$ + 2$K_Q\cdot\log(x_{0}/x)$, with $x_0\sim 0.0012$, $B_0=4.1$ GeV$^{-2}$ and $K_Q=0.14$ GeV$^{-2}$
(very weak $Q^2$ dependence of $B_g$ is neglected).
(The dipole fit to $F_{2g}(\Delta,x)$ gives a very similar numerical result for \effs \, in our kinematics, decreasing \effs  \,
by 4-5$\%$ which is well within the uncertainties of the current knowledge of the t-dependence of the gluon GPD in the studied  $x,Q^2$ range ).
 \par Integrating over $\Delta^2$, we obtain for
  \effs\ in
  the mean field approximation:
\beq
\frac{1}{\sigma^{(MF)}_{\textrm{\scriptsize eff}}}=\frac{1}{2\pi}\frac{1}{B_g(x_1)+B_g(x_2)+B_g(x_3)+B_g(x_4)},\label{mura}
\eeq
where $x_{i}$ are the longitudinal momentum fractions of the four
partons involved  in the \22 mechanism.
Hence we find for the  mean field value of $\sigma_{eff}$\,  in
  the LHCb  kinematics  $x_2\sim x_4=0.02,x_3\sim x_4\sim 0.0001$:
\beq \sigma^{MF}_{eff}\approx 40 mb,\eeq
which, as we already mentioned, is a factor of two larger than  the
the value reported by the LHCb.
\section{3 to 4 mechanism}
\par The mechanism for the enhancement of the rate of DPS (increase of
$1/\sigma_{eff}$ ) as compared to its
 mean field value  was suggested in \cite{BDFS1,BDFS2,BDFS3,BDFS4}, where it was shown that taking into account the pQCD DGLAP ladder splits leads to a decrease of $\sigma_{eff}$\, - the \12 mechanism, see the right hand side of Fig. 2.

We calculate $R$
 by solving by iterations the evolution equation for $_2GPD$ \cite{BDFS2,BDFS4,Gauntnew}.
 The cross section due to the \12 mechanism is calculated as \cite{BDFS2}
\begin{eqnarray}
\frac1{\sigma_{\textrm{\scriptsize eff}}} &\equiv&
\int \frac{d^2\vec{\Delta}}{(2\pi)^2}\mbox{ } _{2}D_2(x_1,x_2, Q_1^2,Q_2^2;\vec\Delta)  \cdot {}_2D_2(x_3,x_4, Q_1^2,Q_2^2; -\vec\Delta)\nonumber\\[10pt]
&+&_2D_1 (x_1,x_2, Q_1^2,Q_2^2;\vec\Delta)
\cdot {}_2D_2(x_3,x_4, Q_1^2,Q_2^2;-\vec\Delta)\nonumber\\[10pt]
 &+& {}_{2}D_2(x_1,x_2, Q_1^2,Q_2^2;\vec\Delta)\cdot {}_{2}D_1(x_3,x_4, Q_1^2,Q_2^2;-\vec\Delta).\label{2}
 \end{eqnarray}
 Note here that the \oo  mechanism contribution must be excluded \cite{BDFS2}.
Here $_2D_1$ corresponds to \12 mechanism, while $_2D_2$ to \22 contribution (with generic initial conditions - either factorized, or including  non-factorized terms).

The distribution $_2D_1$ corresponding to Fig. 2
 (the right hand side)
is obtained by solving by iterations of  the evolution equation for $_2GPD$ \cite{BDFS2,BDFS3,Gauntnew}.  The mean field distribution $_2D_2 $ gets corrections from the QCD evolution due to \12 mechanism.
The DPS effective cross section is then parametrized as
\beq
\sigma_{DPS}=\sigma_{\rm MF}/(1+R_{pQCD}).\label{7}
\eeq
In \cite{BDFS2,BDFS3}  it was assumed that the factorized form  given by
eq. \ref{fac} is valid at  the starting point of the evolution,
$Q_0^2$, which is essentially the parameter separating soft and hard dynamics.
The enhancement coefficient increases with decrease of $Q_0^2$.

 Numerical
 results
  for the enhancement coefficient $R_{pQCD}$ for charm pair production are given in Figs. 6,7 below.
The direct calculations of \effs$_{pQCD}=\sigma_{MF}/(1+R_{pQCD})$ shows that the pQCD correlation leads to a decrease of $\sigma_{\eff}$ to 24--28 mb for the double charm production, slightly larger reduction for charm + bottom, and a more significant reduction for double bottom.
   \par Thus the  \12 mechanism significantly improves the agreement with experimental data, but its relative contribution is smaller than in the central rapidity range  covered by the ATLAS and CMS experiments.

 \section{Non-factorized
 contribution
  to $_2D$ at the initial  $Q_0$ scale.}

 \par
\begin{figure}[h]  
\includegraphics[width=0.48\textwidth]{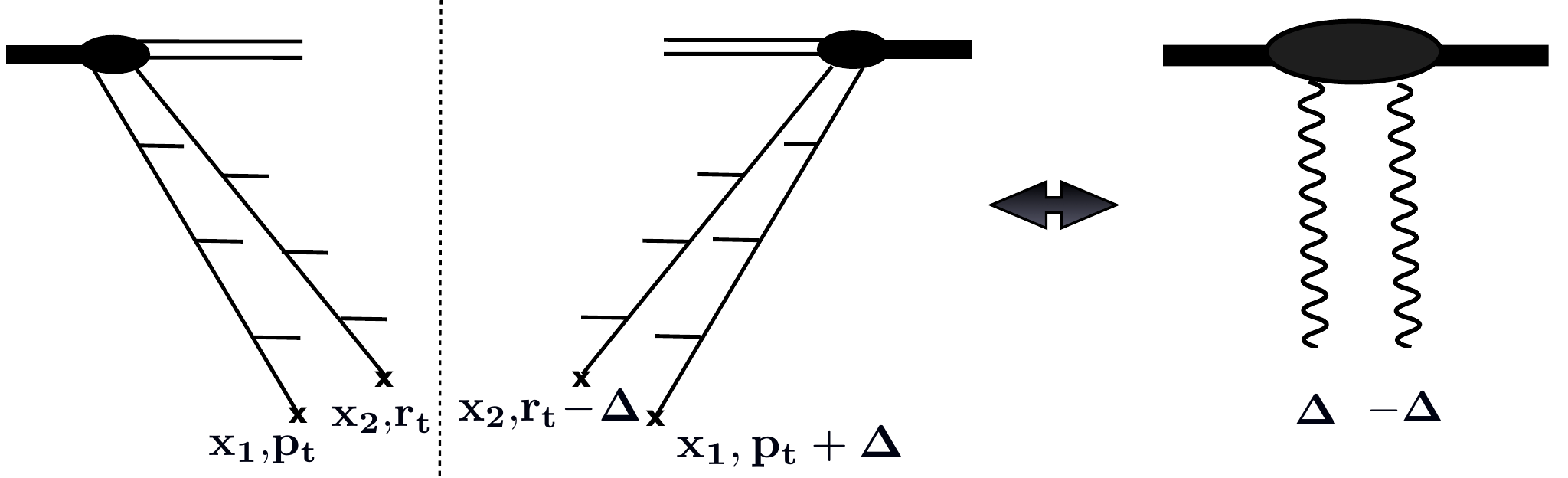}
   \caption{ $_2$GPD as a two Pomeron exchange}
    \label{geom1}
 \end{figure}

 There is an additional contribution to the  DPS at small $x$ which is absent in the case of 
 processes
 involving $x_i \ge 0.01$ (production of jets, etc at the central rapidities).
 This contribution was first discussed in \cite{BDFS3}. It results in
 a non-factorized contribution to $_2$GPD at the initial scale $Q_0^2$ that separates soft and hard physics and which we consider as the starting scale for the DGLAP evolution.
 In the previous sections we assumed that at this scale
 $_2$GPD
 factorizes into the product of two $_1$GPDs.
It is natural to expect that transition from soft to hard QCD regime is smooth
and occurs at
scales $Q^2 \sim 0.5-1\, {\rm GeV}^2$. In this case one expects that at such  a scale
the single parton distributions at small $x$ below $10^{-3}$ are given by the soft Pomeron exchange.
In this picture the two soft partons   may originate from two independent ``multiperipheral ladders'' represented by {\em cut Pomerons}, see Fig.~\ref{geom1}.

\begin{figure}[h]  
  \includegraphics[width=0.48\textwidth]{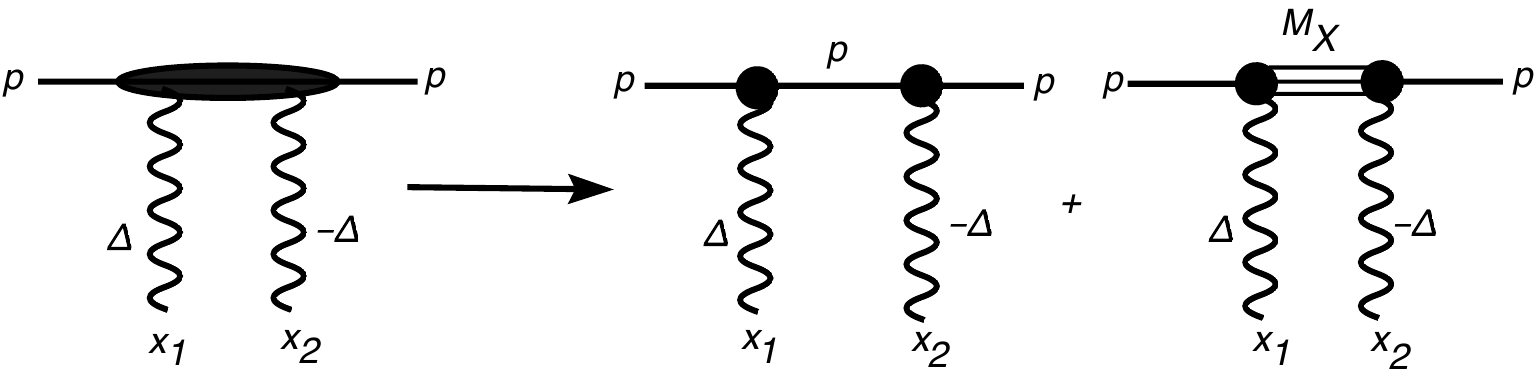}
  \vspace*{-0.3cm}
   \caption{$2\Pomeron$
   contribution to $_2$D and Reggeon diagrams}
    \label{geom2}
 \end{figure}

 The soft Pomeron amplitude is practically pure imaginary \cite{Gribov}
see also \cite{Totem} for most recent experimental measurements.
As a result, this amplitude equals to
 the amplitude of  the {\em diffractive cut}\/ of the two-Pomeron diagram of Fig.~\ref{geom2}.
The  two contributions to the cut are the elastic and diffractive intermediate states.
The elastic intermediate state obviously corresponds to  the uncorrelated contribution to $_2$D, while the inelastic diffractive cut encodes correlations.

\par Note here that in difference from the conventional situation of diffraction into large masses, the  rapidity intervals occupied by  the Pomeron ladders from which partons with fractions $x_1, x_2$ are taken are different.

In the case of soft  diffraction the  ladder corresponding to diffraction to masses $M^2$  occupies the interval of rapidities $\sim \ln(M^2/m_0^2)$ where $m_0^2\sim m^2_N$ is a soft scale. Hence the ladders associated with  the transition $p\to "diffractive\, \, state"$ carry the fraction of the nucleon momentum $x \sim m_0^2/M^2$.

It is convenient to consider first  the ratio of non-factorized
 (correlated) and
factorized
(uncorrelated) contributions   at the zero momentum transfer $t=-\Delta^2=0$:
\begin{equation}
\rho(x_1,x_2, Q_0^2) = {_2D_{nf}(x_1,x_2,Q_0^2)\over D_{f} (x_1,x_2, Q_0^2)} = {_2D_{nf}(x_1,x_2,Q_0^2)\over D_N(x_1,Q_0^2)D_N(x_2,Q_0^2)},
\label{tram1}
\end{equation}

We can write
\begin{equation}
\rho(x_1,x_2, Q_0^2)= \int dM^2 S(M^2)
{D_N(x_1/x,Q_0^2)D_N(x_2/x,Q_0^2)\over D_N(x_1,Q_0^2)D_N(x_2,Q_0^2)},
\label{tram2}
\end{equation}
where factors $x_i/x$ take into account a smaller rapidity intervals occupied by the ladders in the case of transition to inelastic diffractive states. The factor
\beq S(M^2)=C_{3\Pomeron}(M^2/m_0^2)^{\alpha_{\Pomeron}(0) }\label{s}\eeq
corresponds to the cut Pomeron that splits into two Pomerons in diagram 4. It is equal to the product of the triple Pomeron vertex and square of proton - Pomeron residues, cf. \cite{Gribov,Kaidalov}. Here we use $\alpha_{\Pomeron}(0)=1.1$ corresponding to a soft effective Pomeron
\cite{Landshoff}.

\par If $x_1=x_2$ the right hand side of Eq.\ref{tram2} is equal to
\beq
 \int d M^2 {{d\sigma_{\rm in \, diff} (M^2) \over dt dM^2}\over {d \sigma_{\rm el} \over d t}}{\left. \right\vert_{t=0} }
\eeq
that is to  the ratio of inelastic and elastic diffraction in DIS for the invariant $\gamma p$ energy $s=m_0^2/x$.

Using the triple reggeon parametrization of the cross section we can determine
 normalization of the three pomeron vertex $C_{3\Pomeron}$ in eq. \ref{s}   from the HERA data \cite{Aaron:2009xp,H1}
for the   ratio of inelastic and elastic diffraction at $t=0$ in the processes of vector meson production:
\beq
\omega\equiv { {d\sigma_{in.\,  dif.}\over d t}\over {d\sigma_{el}\over d t}}{\left. \right\vert_{t=0} } =0.25 \pm 0.05,\label{cr}
\eeq
The constant $C_{3\Pomeron}$ is   roughly the same for diffractive production of light mesons and $J/\psi$ in a wide range of $Q^2$, thus confirming the hypothesis of a smooth transition between soft and hard regimes. It is determined from the condition
$\rho(x_1,x_1,Q_0^2)=\omega$, where $x_1\sim 0.001$, that corrsponds to HERA data in  \cite{Aaron:2009xp,H1}.
Note here that to have a smooth connection with the low $Q^2$ gluon density model of GRV we take the x-dependence of gluon  density at small $x$ from this model. This may  corresponds to relatively hard effective  Pomeron in the lower legs  though a priori density of partons in the Pomeron may grow more   rapidly  at small $x$ than the overall  Pomeron dominated amplitude.

In the Reggeon calculus \cite{Gribov} the effective triple Pomeron coupling  is expected to decrease slowly with energy due to screening corrections somewhat  reducing the rate of the increase of $\omega$ expected in the unscreened triple Pomeron model.

In any case,  our procedure involves  normalizing parameters of the model for $x\sim 10^{-3}$ and  studying a relatively narrow $x$ range $10^{-4} < x < 10^{-2}$. As a result  our results are not sensitive to the variation of the Pomeron intercept between the soft and hard values.

Accordingly, for the parton  density in the ladder we use:

\beq
xD(x,Q_0^2)=\frac{1-x}{x^{\lambda (Q_0^2)}},
\eeq
where the small $x$ intercept of the parton density $\lambda$ is taken from
 the GRV parametrization \cite{GRV98} for the nucleon gluon  pdf at $Q_0^2$ at small x. Numerically $\lambda(0.5 \,\,{\rm GeV}^2)\sim 0.27$,
 $\lambda(1.0 \,\,{\rm GeV}^2)\sim 0.31$
\par Using eqs. \ref{cr},  \ref{tram2}, \ref{s}  and the above values of $\lambda (Q_0^2)$ we  obtain $C_{3\Pomeron}=0.125\pm 0.025$ GeV$^{-2}$ for $Q_0^2=0.5$ GeV$^2$, and $C_{3\Pomeron}=0.14\pm 0.025$ GeV$^{-2}$ for $Q_0^2=1$ GeV$^2$.

 As a result we can estimate $_2D(x_1,x_2,Q_0^2)_{nf}$ as
\beq
_2D(x_1,x_2,Q_0^2)_{nf}=c_{3\Pomeron}\int^1_{x_m/a} \frac{dx}{ x^{2+\alpha_{\Pomeron}}} D(x_1/x,Q_0^2)D(x_2/x,Q_0^2),\label{d1}
\eeq
where we
 introduced an additional factor  of $a=0.1$ in the limit of integration over $x$ (or, equivalently, the limit of integration over diffraction masses $M^2$) to take into account that the Pomeron exchanges  should occupy  at least two  units in rapidity, i.e.  $x>max(x_1,x_2)/0.1$.
 The dependence on rapidity gap cutoff is weak, of order 10 $\%$, and is present in all inelastic diffraction calculations \cite{Kaidalov}.
 The constant $c_{3\Pomeron}=m_0^2C_{3\Pomeron}$, where  $m_0^2=m_N^2=1$ GeV$^2$ is the low limit of integration over diffraction masses.

\par Consider now the $t= - \Delta^2$ dependence of the above expressions. Strictly speaking
all eqs.~ \ref{tram1},\ref{tram2},\ref{d1}  have to include the explicit dependence on $t$. Here we shall however assume the factorization
of the  t-dependence, that reveals itself in the form
\beq
\frac{d\sigma}{dt}\sim U(x_1,x_2,Q_0^2)F(t),
\eeq
where the function $U$ does not depend on $t$ and all t-dependence is given by the form factor $F(t)$, for which we will use the  exponential parametrization. Then we can use eqs.~\ref{tram1},\ref{tram2},\ref{d1} at $t=0$ (with corresponding functions, given by these equations,
and
 the t- dependence
 given by the exponential form factors $F(t)$). Note that these form factors depend on $x$ and the resolution scale only weakly and the scale dependence  can be neglected while performing integration in eqs. \ref{tram2},\ref{d1}.
Such factorization is known to work well for a pure diffraction case (diagram 4 for $x_1=x_2$, and we expect it to work
in the general case as well.
\par The t-dependence of elastic diffraction is
given by
\beq F(t)=F_{2g}^2(x_1,t)=\exp(B_{\rm el}(x_1)t).
\eeq

Thus  the t dependence of the  factorized contribution to $_2D_f$ is given by
\beq F(t)=F_{2g}(x_1,t)\cdot F_{2g}(x_2,t)=\exp((B_{\rm el}(x_1)+B_{\rm el}(x_2))t/2),\eeq
where $F_{2g}$ is the two gluon nucleon form factor.
\par The t-dependence of the non-factorized term  eq. \ref{d1}
is given by the t-dependence of the inelastic diffraction: $\exp((B_{\rm in}(x_1)+B_{\rm in}(x_2))t/2.)$.
\par
Studies of various diffractive processes, both ``soft'' ($pp\to p +M_X$) and
``hard'' ($\gamma+p\to J/\psi+p$, $\gamma^*+p\to V+p$ with $V=\rho,\omega,\phi, J/\psi$)
indicate that the $t$-dependence of the differential cross section is dominated
by the {\em elastic vertex}\/ $pp\Pomeron \propto \exp(B_{\rm el}t)$ with $B_{\rm el}=5\div 6\> \mbox{GeV}^2$ for $x<10^{-3}$.

Using the  exponential parameterization $\exp(B_{\rm in}t)$ for the t-dependence of the square of the {\em inelastic vertex}\/ $pM_X\!\Pomeron$,
the experimentally measured ratio of the slopes $B_{\rm in}/B_{\rm el}  \simeq 0.28$ \cite{Aaron:2009xp}
translates into the absolute value $B_{\rm in} = 1.4 \div 1.7\, {\rm GeV}^2$.
A much weaker t-dependence of inelastic diffractive residue as compared to the elastic vertex is observed also for reaction  $pp\to p +M_X$, see e.g. \cite{Alberi:1981af}.

In the language of the  Reggeon calculus this is a consequence of the well known observation that the t-dependence of
three Pomeron vertex is much weaker than of the square of the $pp\Pomeron$ vertex,
see e.g.
\cite{Kaidalov}.

\par The evolution of the  initial conditions, eq.~ \ref{d1}, is given then by
 \beq
 _2D(x_1,x_2,Q_1^2,Q_2^2)=\int_{x_1}^{1}\frac{dz_1}{z_1}\int_{x_2}^{1}\frac{dz_2}{z_2}G(x_1/z_1,Q_1^2,Q_0^2)G(x_2/z_2,Q_2^2,Q_0^2) _2D(z_1,z_2,Q_0^2), \label{r}
 \eeq
  where $G(x_1/z_1,Q_1^2,Q_0^2)$ is the conventional DGLAP gluon-gluon kernel \cite{DDT}
  describing evolution from $Q_0^2$ to $Q_1^2,Q_2^2$.
In our calculations we neglect initial sea quark
 densities in the Pomeron at scale $Q_0^2$ (obviously Pomeron does not get contribution from the valence quarks).
\par Let us define the  quantity K (generalizing $\rho$ from eqs. \ref{tram1},\ref{tram2} to arbitrary $Q_1^2,Q_2^2$ ):
\beq
K(x_1,x_2,Q_1^2,Q_2^2,Q_0^2)\equiv \frac{D(x_1,x_2,Q_1^2,Q_2^2,Q_0^2)}{D(x_1,Q_1^2)D(x_2,Q_2^2)}.
\label{g41}
\eeq
The nominator of this quantity is given by integral \ref{r}, while the denominator is a product of the conventional PDFs.
We carried the numerical calculation of $K$ for $Q_0^2=0.5$ GeV$^2$ and
$Q_0^2=1.0 $ GeV$^2$.
 The typical results are  presented in Fig. 5.
(the corresponding $x_i$ are taken in  accordance with analysis of  section 2, and the calculations are carried out at t=0.).

\par One can see that $K$
grows with increase of $Q_0^2$ and that the QCD evolution leads to the suppression of the nonperturbative contribution. We perform calculation neglecting the PPR (Pomeron-Pomeron-Reggeon) contribution. Inclusion of this term would increase the result by $\sim$ 10\%. Overall we estimate the errors in the K-factor due to uncertainties
in the input parameters are   $\sim$ 25-35\%.

The characteristic feature of K-factor is its increase
as one considers
 more forward kinematics for charm production.
  Moreover, if we start from smaller $x_1,x_2$ the rate of decrease of K with the increase of transverse momenta decreases. We illustrate these features in Fig. 5, where
  we consider K for the charm production kinematics described in section 2
  (in Fig.5 $Q_0^2=0.5 $ GeV$^2$, the behaviour for $Q_0^2=1 $ GeV$^2$ is similar. Upper curve is $K$ for the $_2$GPD with small x $\sim 10^{-4}$ gluons and lower one -- for larger x $\sim 10^{-2}$.
  We can see that main non-factorizable contribution originates from a smaller x gluon pair. The same is true for production of $c\bar c b\bar b$ and
  $b\bar b b\bar b$.

\begin{figure}[htbp]
\begin{center}
\includegraphics[scale=0.7]{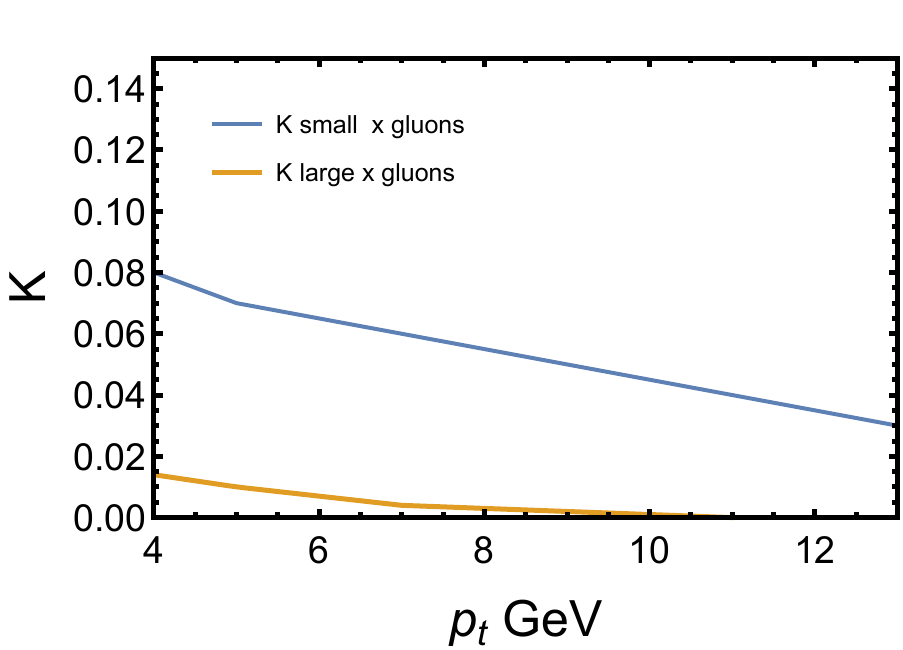}\hspace{3cm}
\label{5b}
\caption{ Transverse
momentum
dependence of K factor \ref{g41} for $_2$GPD  for regimes of small and large x in kinematics of chapter 2 ($Q_0^2=0.5 $ GeV$^2$ )}
\end{center}
\end{figure}
One can see from Fig.5 that $K(x,Q^2)$ decreases strongly with increase of $Q^2$. This reflects the increase of typical $x$ at $Q_0^2$ scale contributing to $K(x,Q^2)$ with increase of $Q^2$ and a fast decrease of $K(x,Q_0^2)$ with increase of $x$ (remember that $K(x\ge 0.05-10^{-1},Q_0^2)\approx 0$ and grows strongly  with decrease of x less than $10^{-2}$.
\section{Contribution of the correlated term in the  initial conditions to DPS.}
\par We can now write the general expression for \effs \, taking into account non-factorized contribution to the  initial conditions, the  \12 mechanism and  the mean field contribution.
\begin{eqnarray}
\frac{1}{\sigma_{\rm eff}}&=&\int \frac{d^2\Delta}{(2\pi)^2}(\exp(-(B_{1\rm el}+B_{2\rm el})\Delta^2/2)+S_{12\, \rm  pQCD}+K_{12}\exp(-(B_{1\rm in}+B_{2 \rm in})\Delta^2/2.))\nonumber\\[10pt]
&\times&(\exp(-(B_{3\,\rm el}+B_{4\,\rm el})\Delta^2/2)+S_{34\, \rm  pQCD}+K_{34}\exp(-(B_{3\rm in}+B_{4\rm  in})\Delta^2/2))).\nonumber\\[10pt]
\label{f1}
\end{eqnarray}
Here $B_i\equiv B(x_i)$, and
\beq S_{ij\, \rm  pQCD}\equiv S(x_i,x_j,Q_i^2,Q_j^2)=\frac{_2D_1(x_i,x_j,Q_i^2,Q_j^2)}{D(x_i,Q_i^2)D(x_j,Q_j^2)}.\eeq
Also
\beq K_{ij}\equiv \frac{_2D(x_i,x_j,Q_i^2,Q_j^2)_{\rm  nf}}{D(x_i,Q_i^2)D(x_j,Q_j^2)},\label{kij}\eeq
is the ratio of $_2$GPD obtained from non-factorized and factorized
terms
 at the scale $Q_1^2,Q_2^2$.
After carrying
out
 integration over $\Delta^2$ we
 obtain the expression for \effs \,  in terms of $R_{\rm  pQCD} $, K, $B_{\rm el}$
  and $B_{\rm  in}$. For simplicity we will  write it only for the
  case of kinematics under considerations
  where the  $K$ term enters only for  the partons with  smaller $x$'s.

\section{\effs \, for production of the heavy quark pairs}.
\par We can now return to the analysis of the process of production of two charmed pairs. We consider the symmetric kinematics, i.e. $x_1\sim x_2; x_3\sim x_4$.

In this case we can neglect terms proportional to $K_{34}$ since it corresponds to a negligible Regge mechanism contributions at $x_3,x_4\sim 0.01\div 0.1$, and in particular neglect $S_{12}$--$K_{34}$ interference terms). Then we have
\beq
\frac{1}{\sigma_{\rm  eff}}=\int \frac{d^2\Delta}{(2\pi)^2}((\exp(-B_{1\rm el}\Delta^2)+K_{12}\exp(-B_{1\rm  in}\Delta^2)+S_{12})(\exp(-B_{3 \rm el}\Delta^2)+S_{34})-S_{12}S_{34}) ,
\eeq
 Carrying out the integration we obtain for the full rescaling of \effs \,  including all three  mechanisms discussed above:
\beq
R_{\rm tot}=R_{\rm pQCD}+R_{\rm soft},
\eeq
where $R_{\rm pQCD}$ is the cross section enhancement due to \12 mechanism, i.e. proportional to $S_{34}exp(-B_{1\rm el}\Delta^2)$,see section 3, while
\beq
R_{soft}=K_{12}(
\frac{B_{1 \rm el}+B_{3 \rm el}}{B_{3 \rm el}+B_{1 \rm in}}+R_{  \rm pQCD}\frac{B_{1  \rm el}}{B_{1 \rm in}}),
\eeq
is the enhancement due to nonperturbative
correlations
and interference of nonperturbative and perturbative contributions.

Note that the main
sources  of large $R_{tot} $
are
 the presence of the pQCD enhancement -- \12 for two partons with larger $x$ and nonperturbative enhancement for smaller x's. The latter
enhancement is amplified by the fact that
the only $\Delta^2$ dependence in this case due to $\exp(-B_{ \rm in}\Delta^2)$, whose slope is almost three times smaller than
that of the mean field term,
leading to the major enhancement of the corresponding contribution, compensating relatively small K.
(The smallness of K is connected with
a rapid decrease
of the effect of non-perturbative correlations  with the increase of $Q^2$.)
 Thus the enhancement we obtain
is essentially due to asymmetric (between upper and lower parts of diagram Fig. 2)
kinematics of two pairs of x's.
\par Numerically , $B_{1 \rm el}+B_{2 \rm  el}\sim
8.2
$ GeV$^{-2}$, $B_{1 \rm el}/B_{1 \rm in}\sim 2.8$.
Thus for example for $p_t=4 GeV$
altogether the Regge type contribution to $R$ is $\sim 0.3$,$R_{ \rm pQCD}\sim 0.7$
For $Q_0^2=1 GeV^2$ we find  the Regge contribution to $R$ to be  larger--$\sim 0.4$,
while $R_{ \rm pQCD}\sim 0.4$, As a result
for both choices of the initial conditions we obtain  $R\sim 1.8-2.$,
leading to
\beq
\sigma_{\rm eff}\sim 20-22 mb \label{list}
\eeq
Note that numerically variation of the  values of $R_{ \rm pQCD}$, with
a choice of
the starting point of the $Q^2$ evolution
is practically
completely compensated by the variation
 of the  soft non-factorizable contribution.  \par Note that eq.\ref{list} does not include additional uncertainties in the Reggeon
 calculation. For example, uncertainty in the ratio
of inelastic and elastic diffraction of order $25\%$ will lead to $19-23$ mb in eq.\ref{list} and so on.
 There is a similar uncertainty due to the input t-dependence of the gluon GPDs.

 \par The same calculation for the production of two bottom and two charm pairs in the LHCb kinematics
\cite{Belyaev,LHCb}
also gives
 $R\sim 1.9-2$. In this case $\sigma_{\rm eff \, mean \, field}\sim 38 mb$, and we find
 \effs$\sim 19$ mb, in a good agreement with the LHCb
 data.
 \par We
 show  different contributions to
 \effs \ enhancement  as a function
 of the transverse momentum of $D $ meson $p_t$ for 3.5 TeV and 6.5 TeV runs in figures 6 and 7.

 \begin{figure}[htbp]
\begin{center}
\includegraphics[scale=0.7]{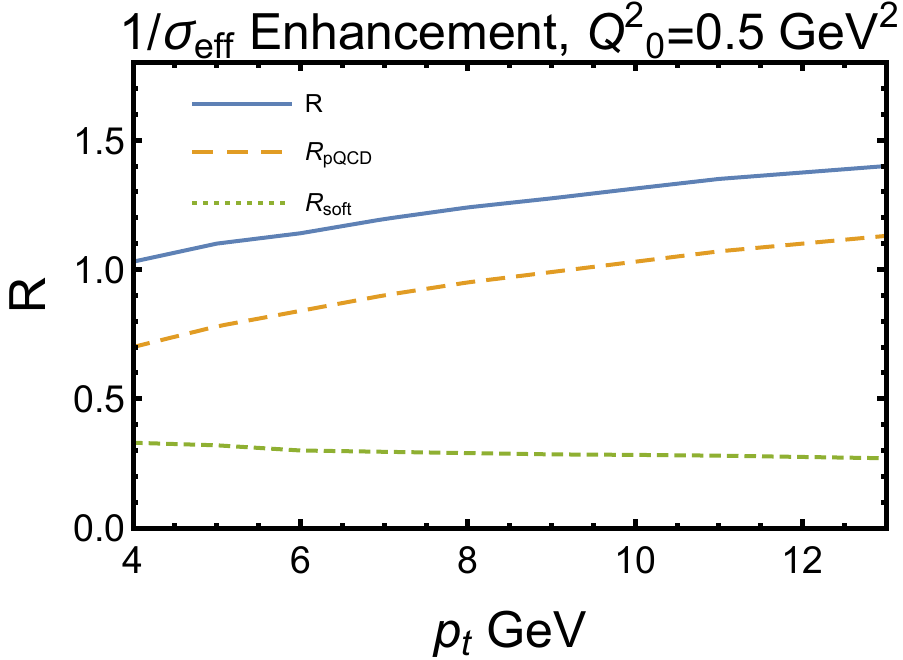}\hspace{3cm}
\includegraphics[scale=0.7]{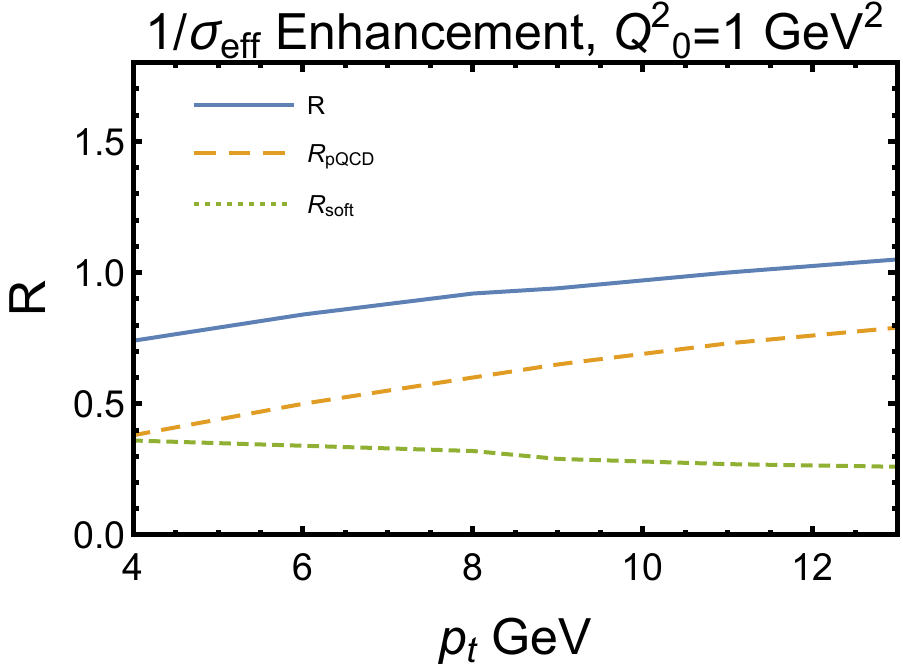}\hspace{3cm}
\label{5b1}
\caption{$R_{tot}$ and contributions to $R_{tot}$ due to
$R_{pQCD},R_{soft}$
as a function of the $D$ meson transverse momentum $p_t$ for 3.5x3.5 TeV run}
\end{center}
\end{figure}

 \begin{figure}[htbp]
\begin{center}
\includegraphics[scale=0.7]{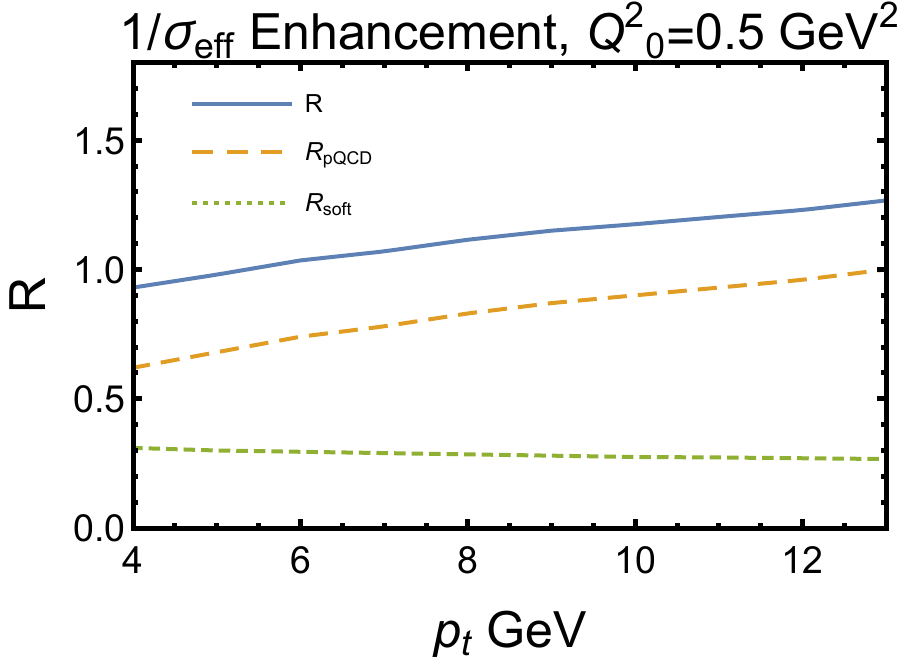}\hspace{3cm}
\includegraphics[scale=0.7]{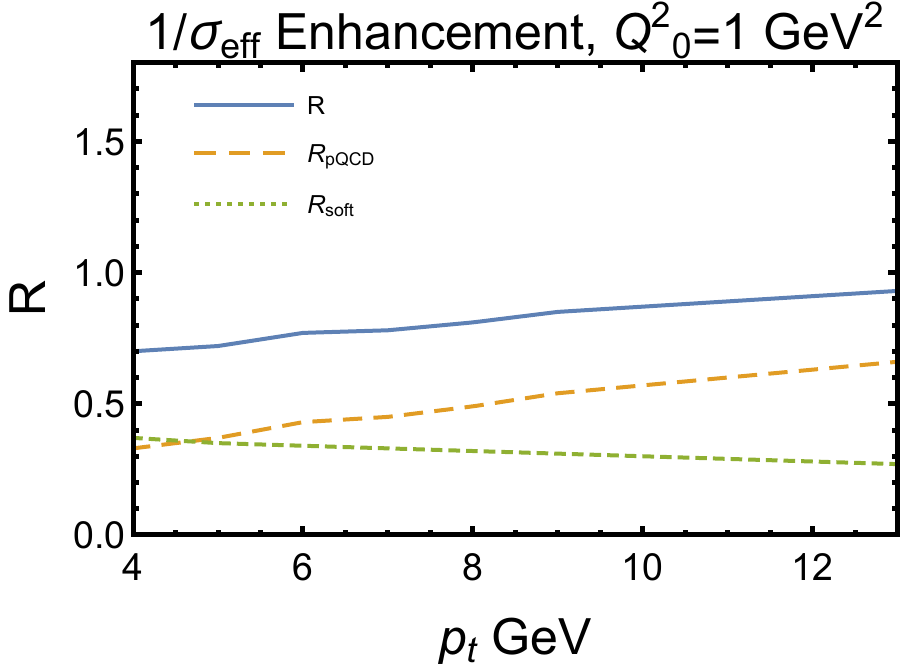}\hspace{3cm}
\label{5b2}
\caption{$R_{tot}$ and contributions to $R_{tot}$ due to
$R_{pQCD},R_{soft}$
as a function of the $D$ meson transverse momentum $p_t$ for 6.5x6.5  TeV run}
\end{center}
\end{figure}
We see that the $R_{pQCD}$ slowly decreases with energy, but this is compensated with increase of $R_{soft}$, whose relative contribution also increases with the increase of energy.
\par The corresponding \effs for
two LHC runs
are depicted in Figs. 8.

 \begin{figure}[htbp]
\begin{center}
\includegraphics[scale=0.7]{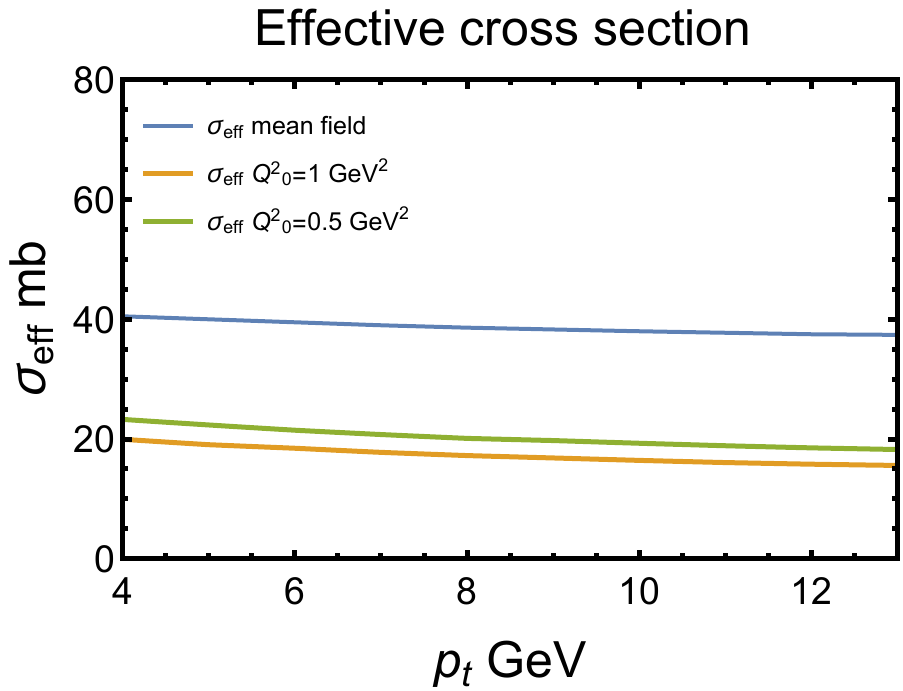}\hspace{3cm}
\includegraphics[scale=0.7]{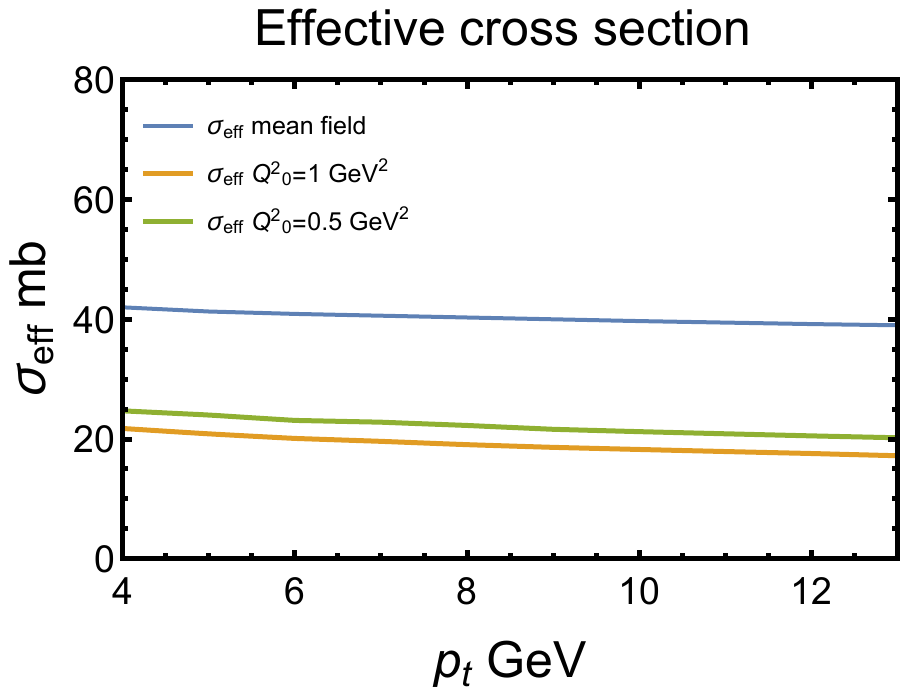}\hspace{3cm}
\label{5b4}
\caption{\effs \,  as a function of the $D$ meson transverse momentum $p_t$ for $Q_0^2=0.5,1$ GeV$^2$ and for 3.5x3.5 TeV and 6.5x6.5 TeV runs}
\end{center}
\end{figure}
\par We see that the \effs increases by less than 1 mb for small $p_t$ 
 when we move from 3.5 to 6.5 TeV,
i.e. it effectively remains constant with the increase of energy,
due to
  increase of soft correlations contribution  compensating the decrease of pQCD contribution and
   increase of mean field $\sigma_{\rm eff}^{MF}$. In fact of course such small changes are beyond the accuracy
of our model, and we can only conclude that \effs are approximately constant in this interval of energies for given transverse momenta $p_t$.
We obtain very similar results
for the production of two pairs of $b\bar b$ (Fig. 9).
Note that in
 our approach the same \effs  are expected for production of two $\Upsilon$ and $\Upsilon b\bar b$, cf discussion in section 2 of the case of charm production.
 \begin{figure}[htbp]
\begin{center}
\includegraphics[scale=0.7]{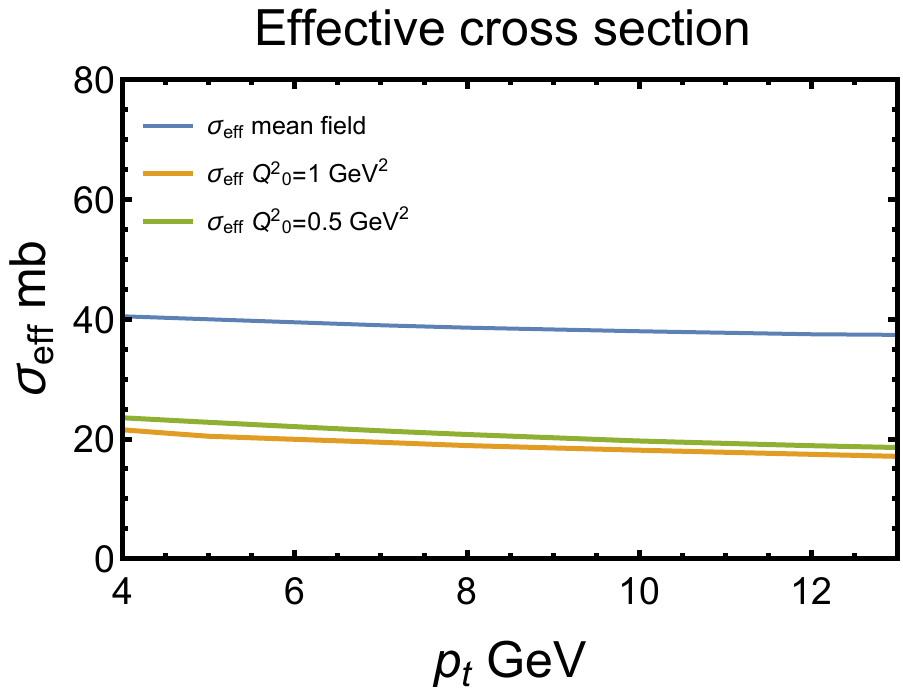}\hspace{3cm}
\includegraphics[scale=0.7]{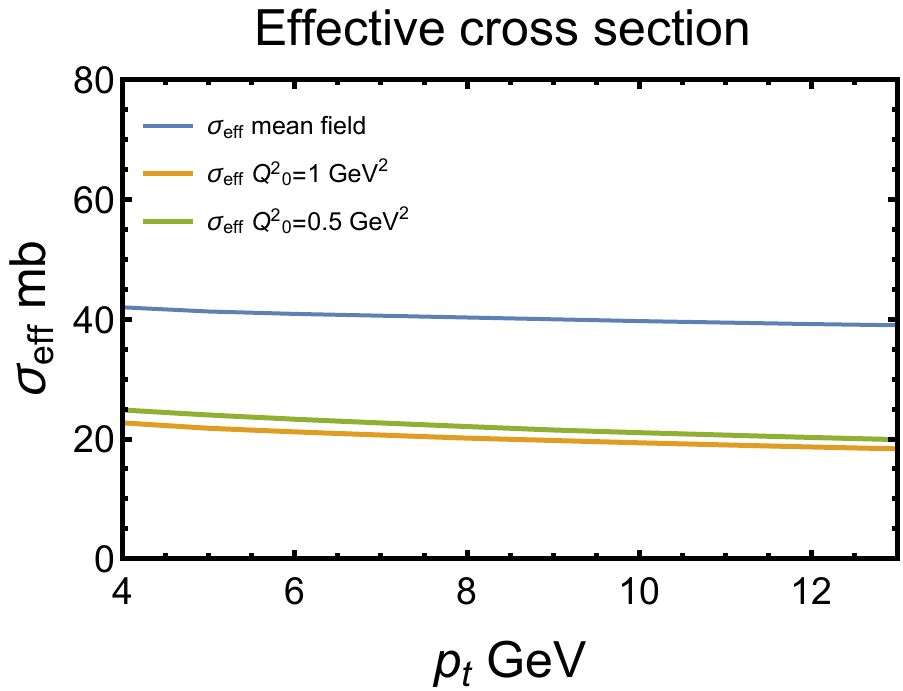}\hspace{3cm}
\label{5b3}
\caption{\effs \,  as a function of the $B$ meson transverse momentum $p_t$ for $Q_0^2=0.5,1$ GeV$^2$ and for 3.5x3.5 TeV and 6.5x6.5 TeV runs}
\end{center}
\end{figure}
\section{Conclusions}
We have demonstrated that the rate of DPS of the  production
of two  pairs of $D$ mesons in the $pp$ collisions in the forward kinematics studied by the
 LHCb can be explained by taking into account two types of correlations in the nucleon double GPD - the  pQCD mechanism of \cite{BDFS1,BDFS2,BDFS3,BDFS4} which allowed previously
 to describe the rate of DPS at the central rapidities and new nonperturbative correlation mechanism specific for small $x$  which is related to the  phenomenon of the inelastic diffraction.
\par  Account for two correlation  mechanisms significantly reduces sensitivity of the results to the starting point of the QCD evolution,
  both for forward and
 for central kinematics.

 Though the estimates of the non-perturbative correlations are only
 semiquantitative, we naturally obtain  $\sigma_{\eff}\sim 20-22 $ mb for the D meson pair production (see Figs. 7,8) which is in a good agreement with experimental data for 3.6 and 4 TeV runs (see Fig.10 in \cite{LHCb}). We obtain similar results for other charm DPS production processes (2 $J/Psi$, $J/\Psi$ and $D\bar D$ pair), and this is indeed observed in experiment \cite{Belyaev} in the forward kinematics (within experimental accuracy).
 For the DPS production of  the bottom quarks we find (see Fig.9)
 \effs $\sim$ 21-23 mb
which is nearly a factor of two smaller than the mean field estimate of
\effs=38 mb.
Thus we observe that combining pQCD correlation mechanism and the  Regge inspired model for initial conditions
we find
 approximately constant \effs \,  of order 20-22 mb for the LHCb kinematics.
\par Our
calculations of  \effs \, were performed  both for the $3.5\times 3.5 \rm \, TeV$
 $6.5 \times 6.5 \rm \, TeV$  runs. (The corresponding differences with 4 and 7 TeV runs respectively are negligible). We obtain practically the same values of \effs
since the decrease  of $R_{\rm pQCD}$ is compensated
by increase if $R_{\rm soft}$. The actual difference is of order 1mb, slightly increasing to 2 mb (\effs slightly decreases with increase of energy, but this change may be artifact of our model assumptions, i.e. it is 
obviously
 beyond the accuracy of our model).

\par Clearly, the role of soft correlations increases with the decrease of typical Bjorken x in the process. The same is true for transverse scale where the soft correlations start to be relevant, we see that it increases with energy. On the other hand the changes in the scale of pQCD and soft correlations tend to compensate each other with the increase of energy.
This means that from the theoretical point of view it will be extremely helpful to carry the measurement of \effs for new 6.5 TeV run at LHCb, as well as to measure dependence of \effs on rapidity of the forward quark pair.
\par
 Obviously
 the calculations of soft correlations  presented in this paper can be considered only as a semi-quantitative estimate.
 In particular this is connected with large uncertainty in the parameters of the model (see section 4) that are known with the accuracy of $25-30\%$, leading to corresponding inaccuracy in $R_{soft}$.
Additional inaccuracy (although significantly reduced) is due to the choice of $Q_0^2$ scale. Nevertheless,
our estimate clearly reveals importance of soft correlations in forward kinematics and increase of their contribution with the energy of the collision.
\par Finally , let us note that our model can be used also for  the central kinematics,
where in particular it can be applied to
calculate
 \effs\,  in the underlying event (UE). Preliminary
results show that it will not influence significantly the MC simulations of UE given in \cite{BG1,BG2},
although it may lead to stabilization of \effs \, in the region of small $p_t$ characteristic for UE.
The detailed results for the  central kinematics, as well as comparison of our predictions for \effs \, with that
of \cite{ost}, that developed a different model also  based on observations of \cite{BDFS3} and applied it to the
central kinematics, will be given elsewhere \cite{MS}.

\section*{Acknowledgements}
\label{Ack}
M.S.'s research was supported by the US Department of Energy Office of Science,
Office of Nuclear Physics under Award No.  DE-FG02-93ER40771.
We thank Ivan Belyaev, Yuri Dokshitzer and Leonid Frankfurt for many useful discussions.
We also thank the TH department of CERN during the time this study was completed.

\end{document}